\documentclass[%
 reprint,
superscriptaddress,
 amsmath,amssymb,
 aps,
]{revtex4-2}

\usepackage[dvipsnames]{xcolor}
\usepackage{graphicx}
\usepackage{dcolumn}
\usepackage{bm}
\usepackage{amsmath}

\usepackage[utf8]{inputenc}
\usepackage[english]{babel}

\DeclareUnicodeCharacter{03BC}{\ensuremath{\mu}}
\makeatletter
\let\oldselectlanguage\selectlanguage
\renewcommand{\selectlanguage}[1]{%
  \def\tempa{#1}%
  \def\tempb{en}%
  \ifx\tempa\tempb
    \oldselectlanguage{english}%
  \else
    \oldselectlanguage{#1}%
  \fi
}
\makeatother

\begin{document}

\preprint{APS/123-QED}

\title{ Microfluidic Oscillatory Rheology of Transported Soft Particles
}

\author{Matteo Milani}

\affiliation{PMMH, ESPCI, Universit\'e PSL, CNRS, Sorbonne Universit\'e, Universit\'e Paris Cit\'e, 75005 Paris, France}

\author{Joshua D. McGraw}
\affiliation{Gulliver, ESPCI, Universit\'e PSL, CNRS, IPGG, 75005 Paris, France}

\author{Anke Lindner}
\affiliation{PMMH, ESPCI, Universit\'e PSL, CNRS, Sorbonne Universit\'e, Universit\'e Paris Cit\'e, 75005 Paris, France}
\affiliation{Institut Universitaire de France (IUF), Paris, France}

\author{Stefano Aime}
\email{stefano.aime@espci.fr}

\affiliation{Molecular, Macromolecular Chemistry, and Materials, ESPCI Paris, 75005 Paris, France.}

\date{\today}
\begin{abstract}

Microfluidic channels have emerged as useful tools to control dynamic forcing on transported microscale objects, as encountered in emulsions, biological flows, and other soft matter systems. Tailored channel designs enable precise interfacial and bulk rheological measurements of complex materials over a wide range of forcing timescales. After a brief overview of recent experiments illustrating these techniques, we discuss perspectives for future research in this direction, including the study of lubrication films in highly confined droplets, the measurement of fast relaxation dynamics of complex interfaces, and the high-throughput rheological characterization of microscopic soft matter systems ranging from single macromolecules to cells.

\end{abstract}

\maketitle


\section{Introduction}

Microfluidics is becoming well-established for performing rheological measurements on complex fluids~\cite{pipe2009microfluidic,lindner2016preface}. 
Compared to conventional rheometry, microfluidic rheology offers several advantages. First, it requires extremely small sample volumes. Second, microfluidic devices access a wide range of shear and extensional flow conditions, including deformation rates beyond those attainable by standard rheometers. Microfluidic devices also facilitate integration with optical microscopy or other in-situ characterization techniques, enabling direct observation of the flowing fluids. 

\begin{figure}[t!]
\centering
\includegraphics[width=1\columnwidth]{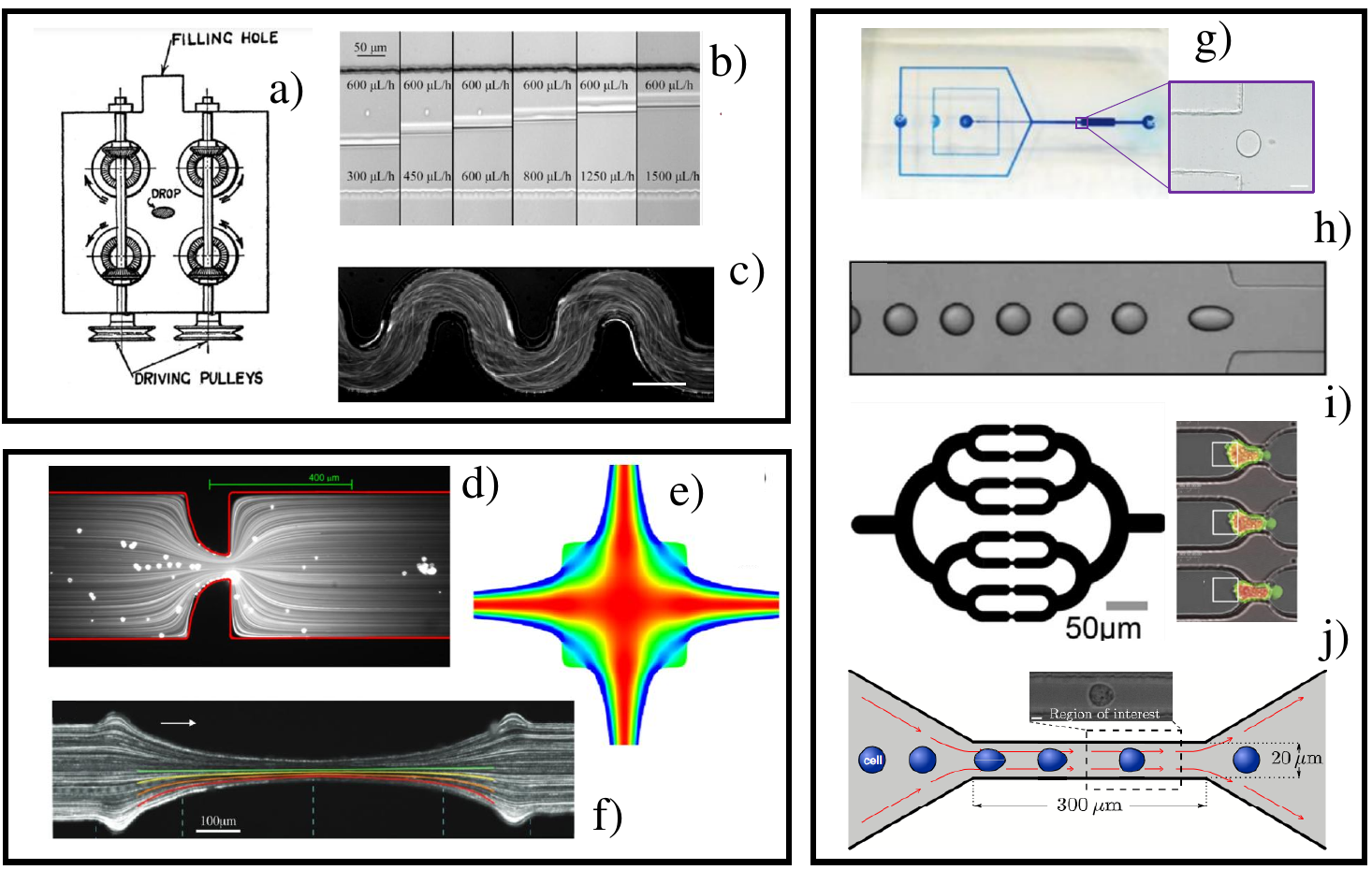}
\caption{\label{fig:modulus_frequency}\textbf{Examples of microfluidics channels used for rheological study:}  \textbf{(a)} Original sketch of the four-roller apparatus reprinted from~\cite{taylor1934formation}. \textbf{(b)} Image of the parallel flow measured for various flow rates in the microfluidic channel, reprinted from~\cite{guillot2006viscosimeter}. \textbf{(c)} Streakline image of unstable flow in a microfluidic serpentine channel, reprinted from~\cite{zilz2012geometric}. \textbf{(d)} Flow pattern in a hyperbolic contraction, reprinted from~\cite{campo2011flow}. 
\textbf{(e)} Strain rate field for numerically simulated
Newtonian creeping flow in the optimized cross slot geometry, reprinted from~\cite{haward2012optimized}.
\textbf{(f)} Flow
patterns in an optimized hyperbolic channel, reprinted from~\cite{liu2020optimised}.
\textbf{(g)} The microfluidic chip used in Ref.~\cite{andre2019new}, the inset shows an oil droplet that undergoes an expansion. 
\textbf{(h)} Sequence of drops deforming progressively in an extensional flow gradient caused by a flow constriction of the channel, reprinted from~\cite{cabral2006microfluidic}. 
\textbf{(i)} Schematic of the microfluidic bypass device. On the right sequential
micrographs of a cell entry into a constriction, reprinted from~\cite{lange2015microconstriction} \textbf{(j)} Sketch of a cell flowing into Channel geometry for RT-DC measurement. The inset representatively
shows a deformed HL60 cell, reprinted from~\cite{mietke2015extracting}.}
\end{figure}

One class of simple microfluidic rheometers is based on determining the relation between applied pressure differences and flow rate, replacing the torque–rotation rate measurements used in conventional rheometry. These devices typically employ straight or hyperbolic channels to measure shear or elongational viscosities~\cite{pipe2009microfluidic}. More sophisticated geometries ---mimicking for example Taylor's four-roller apparatus~\cite{taylor1934formation} shown in Fig.~\ref{fig:modulus_frequency}(a)--- were recently implemented~\cite{hudson2004microfluidic} to probe extensional rheology. 

The ability to design tailored channel geometries in combination with direct visual observation of a flowing complex fluid allows for probing a wide range of rheological properties and operational regimes. For instance, momentum balance at the interface between two co-flowing fluids in a straight channel downstream of a Y-junction (see Fig.~\ref{fig:modulus_frequency}(b)) can be used to measure several rheological properties. Examples include relative viscosities~\cite{guillot2006viscosimeter} and dynamic surface tensions~\cite{steegmans2009dynamic}, often with improved resolution for low-viscosity fluids~\cite{gachelin2013non} compared to conventional bulk rheometry techniques. In addition, relaxation times of viscoelastic fluids can be determined from the onset of elastic flow instabilities, seen for example in Fig.~\ref{fig:modulus_frequency}(c), triggered by normal stress differences in curved serpentine channels~\cite{zilz2014serpentine}. Another method for relaxation-time assessment is to exploit channel deformability in soft microfluidic devices~\cite{chun2025experimental}. Measuring birefringence in flowing polymer solutions furthermore opens the possibility to locally access viscous stresses without the need of a pressure sensor \cite{haward2012optimized}. 

Furthering the idea that microfluidic geometries can be easily tailored, an exploration of different channel shapes to obtain better-defined stress profiles appeared in the literature over the last decade or two. Extensional flows, for example, have been routinely generated using simple hyperbolic contractions~\cite{campo2011flow}. It was later realised that such flows could be more finely controlled in optimised cross-slot geometries~\cite{haward2012optimized}. Channels with a more complex shape were also designed to produce a purely extensional stress field~\cite{liu2020optimised}, as shown in Figs.~\ref{fig:modulus_frequency}(d-f).

Microfluidics has also been extensively employed for the characterisation of dispersed flowing objects. For example, droplets flowing out of a sudden channel expansion exhibit a sudden deformation~\cite{tregouet_transient_2018} that can be leveraged to measure viscosity~\cite{andre2019new} and interfacial tension~\cite{brosseau_microfluidic_2014}, as indicated in Figs.~\ref{fig:modulus_frequency}(g-h).
Microfluidic techniques have also been applied to more complex systems, including polymeric microcapsules~\cite{xie_interfacial_2017, tregouet_microfluidic_2019}, biomimetic prototissues~\cite{layachi2022rheology}, as well as entire cells~\cite{lange2015microconstriction, mietke2015extracting}, as seen in Figs.~\ref{fig:modulus_frequency}(i-j).

Within this dispersed-object framework, comparatively little attention has been devoted to optimising the stress profiles exerted on flowing objects. Only a limited number of studies have addressed the design of microfluidic channel geometries capable of generating well-controlled, homogeneous stress fields. For these geometries, the coupling between transport and flow characteristics can be exploited to tailor residence times or Hencky strains \cite{liu2020optimised}. However, temporal control of the applied stress is equally crucial for probing the dynamic mechanical response of transported particles and for advancing these approaches toward more quantitative rheological characterisation of the dispersed objects. Despite this importance, time-dependent stress fields have only recently been explored. In particular, time-dependent deformation of droplets subjected to a sinusoidal extensional hydrodynamic stress—achieved through a rational design of the microfluidic channel geometry~\cite{milani2026rheofluidics}—has recently been implemented. This new technique, which for the purposes of this article we call ``rheofluidics,'' opens exciting perspectives for probing the oscillatory rheology of a wide range of soft particles transported within microfluidic channels.

\section{Oscillatory droplet rheology in shape optimized channels}

\subsection{Time-dependent Oscillatory stress}\label{SS:OscillatoryStress}

\begin{figure}[b!]
\centering
\includegraphics[width=1\columnwidth]{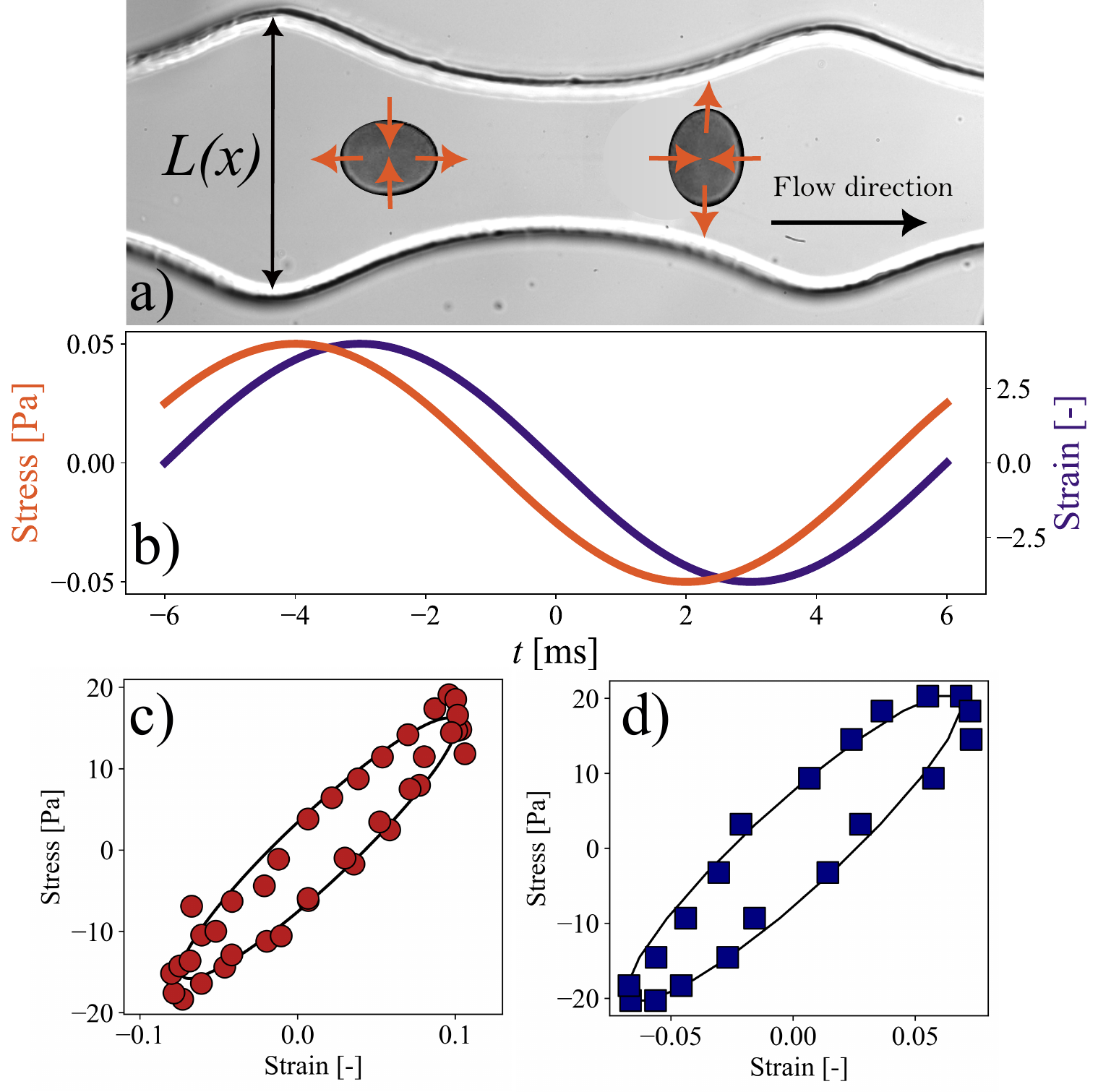}
\caption{\label{fig:example_Rheoflu_Constriction}\textbf{Rheofluidics experiment: }\textbf{(a)} Example of channel section, obtained by solving Eq.~\ref{eq:chshape} with $\sigma_0=50$~Pa, $\omega=500$~rad/s, $q=10$~mm/s, $\eta=10$~mPa$\cdot$s, and $L_0=100~\mu$m. Two snapshots of an oil droplet at maximum and minimum deformation are superimposed; the deformation has been exaggerated to enhance visibility. \textbf{(b)} Sketch of the extensional stress (orange solid line) and the corresponding strain (purple solid line) as functions of time experienced by a flowing droplet. \textbf{(c)} Lissajous plot obtained from the surface rheology of an oil droplet. \textbf{(d)} Lissajous plot obtained from the bulk rheology of a alginate bead.}
\end{figure}

The core idea of rheofluidics consists in applying a well-defined, time-dependent hydrodynamic stress to small samples dispersed in a flowing fluid. This stress control is achieved by guiding the flow through microfluidic channels with a spatially-modulated width, $L(x)$, designed to impose a planar extensional flow field $\dot\varepsilon$ along the channel axis. If the continuous phase is Newtonian with viscosity $\eta$, the planar flow translates into a planar extensional stress $\sigma=\eta \dot{\varepsilon}$. 

As detailed in~\cite{milani2026rheofluidics}, rheofluidics works in the Lagrangian reference frame of particles flowing in the middle of the channel, at speed $v_x(x)=q/L(x)$, with $q$ the planar flow rate. In this reference frame, spatial variations of $L(x)$ translate into a time-dependent stress, $\sigma\left(t(x)\right)$, with $t(x)=\int_0^x ds/v_x(s)$. Thus, $\sigma(t)$ can be prescribed by carefully designing the channel shape, according to the integro-differential equation:
\begin{equation}
\frac{dL}{dx}=\frac{L^2}{q\eta}\sigma\left(t(x)\right)\,.
\label{eq:chshape_general}
\end{equation}

One advantage of this framework is that $\sigma(t)$ can be chosen to suit the needs of the experimenter. The particular case of oscillatory rheology can be obtained in a microfluidic channel whose shape obeys:
\begin{equation}
\frac{dL}{dx}=L^2 \frac{\sigma_0}{q\eta} \sin\left[\frac{\omega}{q} \int_0^x L(s) ds\right]\,,
\label{eq:chshape}
\end{equation}
where $\sigma_0$ and $\omega$ are the amplitude and frequency of stress oscillations, respectively. A typical example of the resulting design, is shown in Fig.~\ref{fig:example_Rheoflu_Constriction}(a). The general shape conforms well to physical intuition: the converging section generates an extensional flow that compresses the droplet vertically, while the expansion induces the opposite deformation. A sequence of well-defined expansions and contractions produces the desired periodic oscillation.
This enables oscillatory rheology in a microfluidic channel, along with a considerable margin for designing other stress profiles. .

\subsection{Experiments}

A typical rheofluidics experiment consists in passing small droplets through the previously-described channel, and measuring their time-dependent deformation. One characteristic deformation mode can be defined as:
\begin{equation}
    \gamma=\frac{d_x - d_y }{d_x + d_y}\,,
    \label{eq:strain}
\end{equation}
where $d_x$ and $d_y$ denote the (time-dependent) sizes of the droplet along the channel axis and perpendicular to it, respectively~\cite{taylor1934formation}. In response to an oscillatory stress shown on the left axis of Fig.~\ref{fig:example_Rheoflu_Constriction}(b), the deformation $\gamma(t)$ typically exhibits oscillations at the same frequency $\omega$ with a phase delay suggesting a viscoelastic response. Typical corresponding data is as sketched on the right axis of Fig.~\ref{fig:example_Rheoflu_Constriction}(b). From such dual data sets, elliptical Lissajous figures can be used to extract the elastic and loss moduli ($G'$ and $G''$), as done by conventional rheology~\cite{Rheoflupy}. 

Earlier, two of us demonstrated the utility of this rheofluidics technique by applying the analysis sketched above to oil droplets and hydrogel beads, for which the different Lissajous figures, illustrated in Fig.~\ref{fig:example_Rheoflu_Constriction}(c-d), reflect distinct material storage and loss mechanisms. In the case of oil droplets, the elastic response is governed by interfacial tension between the two fluids, while the loss modulus is controlled by the extensional viscosity of the continuous phase. By contrast, for hydrogel beads, the response is dominated by the bulk viscoelasticity of the gel network.

\section{Future directions for oscillatory rheology of soft particles in shape-optimized microchannels}

The two examples discussed above demonstrate the strength of such a microfluidic approach for measuring the frequency-dependent rheology of a broad range of microscopic soft matter systems. These measurements rely only on standard microfluidic tools, such as a syringe pump and an optical microscope equipped with a camera, making rheofluidics a promising method for applications across soft matter and biology. In the following, we outline several research directions in which rheofluidics could play a central role, and we discuss technical developments that may further expand its capabilities.

\subsection{Soft particle rheology}

\begin{figure}[b!]
\centering
\includegraphics[width=0.8\columnwidth]{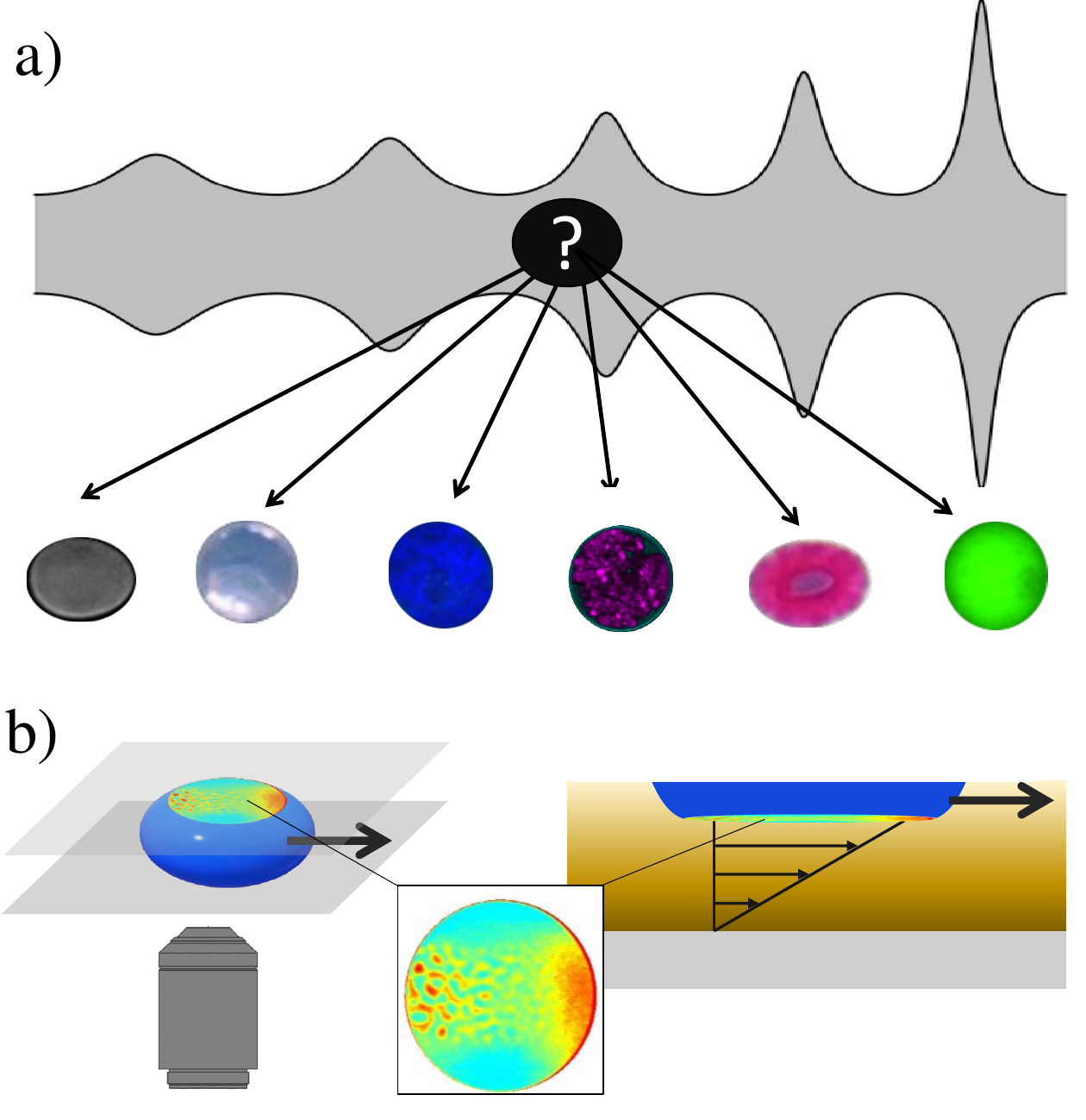}
\caption{\label{fig:finite_size} \textbf{Finite size and interfacial rheology: }\textbf{(a)} Top: sketch of a channel design with constant $\sigma_0$ and decreasing $\omega$.  Bottom: different samples can be probed with such device. From left to right: oil droplet and colloidal gel bead, reprinted from~\cite{milani2025synthesis}; 
rat fibroblast, reprinted from~\cite{fujimoto2008lipid}; bundles encapsulated in terpolymer-stabilized coacervates, reprinted from~\cite{novosedlik2025cytoskeleton}; red blood cell, reprinted from~\cite{marenkov2017annotated}; undifferentiated cells, reprinted from~\cite{olins2004cytoskeletal}; \textbf{(b)} 3D and side-view sketches of a Hele-Shaw droplet flowing through a narrow channel. The compressed region is subjected to shear stress within the lubrication layer. In the 3D view, the droplet is observed from below using a microscope objective, adapted from~\cite{huerre2015droplets}.}
\end{figure}


\textbf{Biological samples:} Like many microfluidic techniques, rheology in shape-optimised channels enables high-throughput measurements. Therefore, hundreds or thousands of independent objects can be probed in as little as a minute using commercial equipment. As such, the technique is particularly suitable for applications requiring the statistical characterisation of large populations of soft objects. A selection of such objects are shown for example in Fig.~\ref{fig:finite_size}(a). Statistical or large-population assessment is needed for most biological systems. Indeed, cell-to-cell variability, rare mechanical responses, and population-level correlations may carry essential physical or biological information. 

In this context, rheofluidics is a promising technique to measure the mechanical properties of cells (including red blood cells, leukocytes, platelets, tumor and stem cells). By probing cellular mechanics under controlled extensional flows, this approach can provide insights into physiological and pathological states, including changes associated with disease~\cite{rianna2016cell}, activation~\cite{janmey2007cell}, or differentiation~\cite{titushkin2007modulation}. Rheofluidics can also be extended to probe a broader range of biological soft objects, including coacervates~\cite{novosedlik2025cytoskeleton}, fibroblasts~\cite{fujimoto2008lipid}, and protein condensates~\cite{charmet2018microfluidics}. These systems are inherently out of equilibrium and exhibit complex viscoelastic behaviours that are closely linked to their biological function \cite{shen_FUS_2023}. Accessing their mechanical properties at the single-object level, while retaining statistical significance, would provide new insights into processes such as intracellular organisation, mechanotransduction, and phase separation in living systems.

\textbf{Gelation and aging of soft materials:} Rapid measurement and fine temporal resolution are also particularly useful for the study of time-dependent phenomena such as gelation and aging. These latter phenomena are relevant for polymerizing systems~\cite{kannurpatti1998study}, colloidal gels~\cite{milani2026decoupling}, phase-separating mixtures~\cite{zhu2001morphological}, glasses and aging emulsions~\cite{cipelletti2003universal}. For this class of systems, shape-optimized channel rheology may complement conventional studies by allowing the investigations of small-scale samples, for which the role of interfaces can be enhanced ~\cite{shen_FUS_2023}. Additionally, channel-based rheology is able to extend measurements to much larger frequencies and give increased sensitivity \cite{milani2026rheofluidics}. Both of these advantages can be important to characterise the earliest stages of the formation of soft gels \cite{keshavarz_timeconnectivity_2021}. 
In addition, the ability to probe gelling systems both mechanically and optically would allow one to directly correlate the temporal evolution of the material’s microstructure with its mechanical response, providing insight into the interplay between dynamics, structure, and rheology in these complex soft materials~\cite{aime_powerlaw_2018}.

\subsection{Lubrication and confinement}\label{LubricationFilms}

Up to now, we have focused on the mechanical properties of unconfined flowing objects such that, using the modelling described in Section~\ref{SS:OscillatoryStress}, it can be assumed that the imposed extensional stress is sinusoidal in time. However, there are many situations in which confinement effects induce considerable dissipation. In these confinement situations, the modelling leading to Eq.~\ref{eq:chshape_general} would need to be considerably revised, representing one opportunity for future theoretical work. On the experimental side, we discuss how this refinement could be assessed in the next section, see ``Local measurements'' there. Presently, we describe the importance of confinement-induced rheology and its study using shape-optimised microfluidic channels. 

\textbf{Droplets and foams:} When the volume of a droplet or bubble is larger than the cube of the channel height, it could be considered as geometrically confined into a Hele-Shaw geometry. This confinement and a motion of the droplets is usually accompanied by the formation of dynamic lubrication layers~\cite{daniel2017oleoplaning, huerre2015droplets}, as indicated in Fig.~\ref{fig:finite_size}(b), between the confining channel walls and the liquid-liquid interface. In particular, the velocity of the droplets determines the lubrication layer thickness~\cite{huerre2015droplets, cantat_liquid_2013, Hodges2004}, which is selected through a balance of viscous and interfacial (\emph{i.e.} surface tension) forces. These menisci are also important in foams~\cite{cantat_liquid_2013, le_merrer_linear_2015} and may depend sensitively on the physical chemistry of the employed surfactants, whose dynamical impacts may be considerable. 

In the context of rheofluidics, the visco-capillary balance noted above contains both dissipative and conservative components. Therefore, an out-of-phase temporal deformation profile of a droplet should be considerably influenced by the friction between a droplet and its confining walls. Moreover, dynamical effects on confined-droplet lubrication layers have hardly been investigated, the majority of the studies known to us having been done in steady state. As schematically illustrated in Fig.~\ref{fig:finite_size}(b), shape-optimised channels could be exploited to investigate dynamic dissipation mechanisms associated with sliding droplets, and thin-film drainage~\cite{steinhaus2006droplet} more generally. 

\textbf{Confined complex fluids:} Many of the droplet studies noted above were concerned with Newtonian fluids. However, additional friction mechanisms such as interfacial slip may be operative in droplets composed of polymer solutions~\cite{ramachandran2012effect, Grzelka2021, Guyard2021}, and may have a significant impact on the throughput and dynamics of confined, complex suspensions. Besides polymer solutions, jammed microgel suspensions are also known to slip significantly~\cite{Meeker2004PRL, Cloitre2017, Pemeja2019}, and would engender an additional storage component, thanks to gel elasticity, in an eventual oscillatory sollicitation. More generally, the interplay between bulk viscoelastoplasticity and surface tension in droplets made of yield stress fluids is a field of active research, as described in another forthcoming contribution to the present collection~\cite{Jalaal_forthcoming}. Without going further into detail, combining confinement and complex fluids engenders the possibility for many types of additional storage and loss components, both bulk and interfacial, not present for unconfined and non-Newtonian situations. It thus appears that rheofluidics offers an ideal method to systematically investigate such complex scenarios, which may be operative in many dynamical and microscale, soft-matter contexts~\cite{perkin2013soft, McGraw2025}.


\subsection{Instrumental perspective}\label{InstPers}

\begin{figure} [b!]
\centering
\includegraphics[width=0.9\columnwidth]{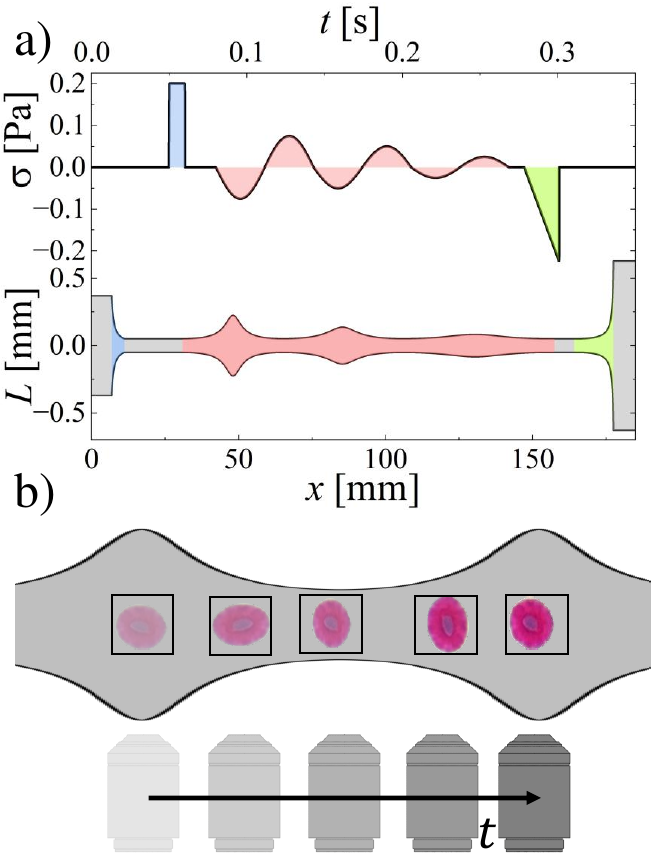}
\caption{\label{fig:instrumental perspective} \textbf{Instrumental perspectives: }\textbf{(a)}Top: sketch of the time-dependent stress profile illustrating a rheological protocol composed of several measurements in series: pre-extension (blue), oscillatory annealing (red), and stress ramp (green) \cite{edera_memory_2025}. Bottom: the corresponding microfluidic channel shape.  \textbf{(b)} Sketch of Lagrangian tracking: the channel is represented as moving backward to account for the stage motion, thereby maintaining the view in the reference frame of the sample.}
\end{figure}


\textbf{Custom stress profiles:} Heretofore, rheofluidics was only used to apply oscillatory stresses. Its core operating principle, expressed by Eq.~\ref{eq:chshape_general}, applies to a much broader range of stress-controlled rheological protocols. For instance, the channel shape associated to a creep test under constant stress $\sigma$ has a relatively simple analytic expression:
\begin{equation}
L_{\mathrm{creep}}= \frac{L_0}{1+\frac{\sigma L_0 x}{q\eta}}\,.
\label{eq:chshape_creep}
\end{equation}
If more complex stress profiles are desired, Eq.~\ref{eq:chshape_general} is more likely to be solved numerically. Interestingly, different solutions can be concatenated to obtain channels with more complex shape, corresponding to elaborated stress profiles mimicking many realistic rheological protocols. A representative example of a protocol including preshear at constant stress, oscillatory annealing at decreasing stress amplitude and a stress ramp is shown together with the corresponding rheofluidic channel shape in Fig.~\ref{fig:instrumental perspective}(a). Similar protocols are often used to study the aging of soft, jammed materials~\cite{edera_memory_2025} that were noted in the previous confinement section.

\textbf{Lagrangian tracking:} rheofluidics experiments have so far been performed in the Eulerian (laboratory) frame. For these experiments, several droplets are observed as they flow through a fixed section of the channel, typically corresponding to a few stress oscillations at most, as shown in Fig.~\ref{fig:example_Rheoflu_Constriction}. To reconstruct the effect of more elaborate stress profiles, such as frequency sweeps, one needs to take several videos, each one observing a different channel section, and analyzing each video to obtain average frequency-dependent properties. While this approach is convenient to probe average properties, it does not allow to measure the full viscoelastic spectrum of one single, individual object. 
To achieve this, one would need to implement Lagrangian tracking~\cite{darnige2017lagrangian,darnige2026deep,junot2022run}, by translating the microfluidic device such that the target flowing object remains in the imaged field of view throughout its flow \cite{liu2020optimised}, as illustrated in Fig.~\ref{fig:instrumental perspective}(b).

\textbf{Local measurements of the stress:} current implementations of rheofluidics rely on the assumption that the stress acting on the flowing droplet corresponds to the one prescribed by channel design. Calibration experiments suggest that this assumption holds to a good approximation if droplets are much smaller than the microfluidic channels~\cite{milani2026rheofluidics}. However, the assumption fails for the highly confined droplets described, for example, in Section~\ref{LubricationFilms} above. In this latter case, or in intermediate cases of moderate confinement, it may be critical to obtain a direct measurement of the hydrodynamic stress tensor around the droplets, as could be done by tracking the flow of the continuous phase by PIV. Such tracking allows for direct visualization of the stress tensor in the channel by assessment of the flow gradients. In addition, indirect information of the local stress may be obtained by integrating into the microfluidic system local pressure sensors based on electric methods~\cite{li2010micro}, conductive inflating membranes~\cite{wang2009polydimethylsiloxane} or capacitive sensors based on composite foam materials~\cite{gauthier2021new}, optical gratings~\cite{hosokawa2002polydimethylsiloxane}, or color changing hydrogels~\cite{ducloue2024color}. 

\textbf{Optics:} rheofluidics requires a method to measure the time-dependent deformation of the flowing droplets. So far, we have done this using bright field microscopy. While this enables the accurate measurement of the droplet cross-section in the middle of the rheofluidic channel, the full, three-dimensional deformation of the droplet remains elusive, and requires assumptions. Using advanced optical techniques such as optical coherence tomography~\cite{huang1991optical,jalaal2018gel}, confocal imaging~\cite{quint20173d}, or holography~\cite{pirone2022speeding} to improve the droplet deformation measurement would be a most valuable improvement. Additionally, micro-optic interferometric systems integrated directly on-chip enable quantitative holographic and phase imaging by measuring light phase shifts at the sample level, eliminating the need for external optical components and enabling compact, stable platforms with faster, real-time access to quantitative data, improved reproducibility, and higher measurement throughput~\cite{bianco2017endowing}.

Using different optical approaches could also be a strategic way to extend the functionality of the technique. For instance, droplet sorting could benefit from laser-based detection methods, which have been successful in microfluidic applications and represent a promising avenue for rheofluidics. In particular, light scattering has been implemented in microfluidic devices~\cite{destremaut2009microfluidics} as a characterization tool, while diffraction-based approaches have also shown promising capabilities for optical interrogation and classification~\cite{masui2025precise}.

\textbf{Data processing with AI: } Recently, many analysis tools are being developed in the fields of rheology~\cite{mahmoudabadbozchelou2024unbiased} and soft matter~\cite{jung2025roadmap,cichos2020machine}. Rheofluidics could serve as an ideal testbed for AI-powered analysis. Its inherently high-throughput nature enables the generation of large datasets suitable for training neural networks, potentially allowing automatic identification of deformation modes, nonlinear responses, and population-level mechanical phenotypes of droplets, beads, and cells under extensional stress. To this end, combining rheofluidics with real-time imaging and machine learning~\cite{guo2026real} could provide unprecedented insight into the mechanics of complex soft materials.




\section{Conclusion}

In summary, shape-optimized microfluidic rheology, and in particular its implementation for oscillatory rheology on chip, is a versatile technique for probing the mechanics of biphasic soft matter at the microscale. The methodology combines precise flow control, direct visualization, and high-throughput. Beyond its initial implementations, further potential lies in the rational design of spatiotemporal stress fields, enabling fully programmable rheological protocols within microfluidic channels. This opens exciting perspectives for accessing previously unexplored regimes, from fast interfacial dynamics and nonlinear responses to the mechanics of highly confined or out-of-equilibrium systems. Coupled with advances in microfabrication, imaging, and data-driven analysis, oscillatory microfluidic rheology could evolve into a quantitative, single-object assessment tool, bridging the gap between bulk measurements and microscopic dynamics, with broad implications for soft matter physics, biology, and materials science. 
\begin{acknowledgments}
The authors benefit from the financial support of PSL University's \emph{Grand Programme: Institut Pierre-Gilles de Gennes}. They likewise gratefully acknowledge the ongoing technical contribution of the joint service unit IPGG Technological Platform CNRS UAR 3750. JDM also thanks the European Union for financial support under the NoDiCE Grant (ERC-2024-COG-101170653). Views and opinions expressed are however those of the author(s) only and do not necessarily reflect those of the European Union or the European Research Council Executive Agency. Neither the European Union nor the granting authority can be held responsible for them. 
\end{acknowledgments}

\bibliography{apssamp}

\end{document}